\documentclass[useAMS,usenatbib]{mn2e}

\usepackage{natbib}
\usepackage{graphicx}
\usepackage{amssymb}
\usepackage{color}
\usepackage{caption}
\usepackage{lipsum,graphicx,multicol}
\usepackage{float}

\usepackage{subcaption}

\title[A new probe of BH spin?]{Radiation pattern and outflow geometry: a new probe of black hole spin? } 
\author[ ]
{W. Ishibashi$^{1,2}$\thanks{E-mail: wako.ishibashi@physik.uzh.ch}, A. C. Fabian$^{1}$ and C. S. Reynolds$^{1}$
\footnotemark[0]\\
$^{1}$Institute of Astronomy, Madingley Road, Cambridge CB3 0HA \\
$^{2}$Physik-Institut, Universitat Zurich, Winterthurerstrasse 190, 8057 Zurich, Switzerland 
}

\begin{document}

\pdfminorversion=4

\date{Accepted ? Received ?; in original form ? \\}

\pagerange{\pageref{firstpage}--\pageref{lastpage}} \pubyear{2012}

\maketitle

\label{firstpage}

\begin{abstract}
We explore the impact of the central black hole (BH) spin on the large-scale properties of the host galaxy, by considering radiative feedback. The BH spin determines the radiation pattern from the accretion disc, which directly imprints on the geometry of the radiation-driven outflows. We show that for low BH spins, the emission is vertically focused, giving rise to polar/prolate outflows; while for high BH spins, the radiation pattern is more isotropic, leading to quasi-spherical/oblate outflows. Reversing the argument, we can potentially deduce the spin of the central BH from the observed morphology of galactic outflows. In principle, this may provide a novel way of constraining the central BH spin from galaxy-scale observations. Indeed, the BH spin can have significant macroscopic effects on galactic scales, ultimately shaping  the large-scale feedback and the resulting obscuration. 
\end{abstract}

\begin{keywords}
black hole physics - galaxies: active - galaxies: evolution  
\end{keywords}


\section{Introduction}

\begin{figure*}
\begin{multicols}{3}
    \includegraphics[width=\linewidth]{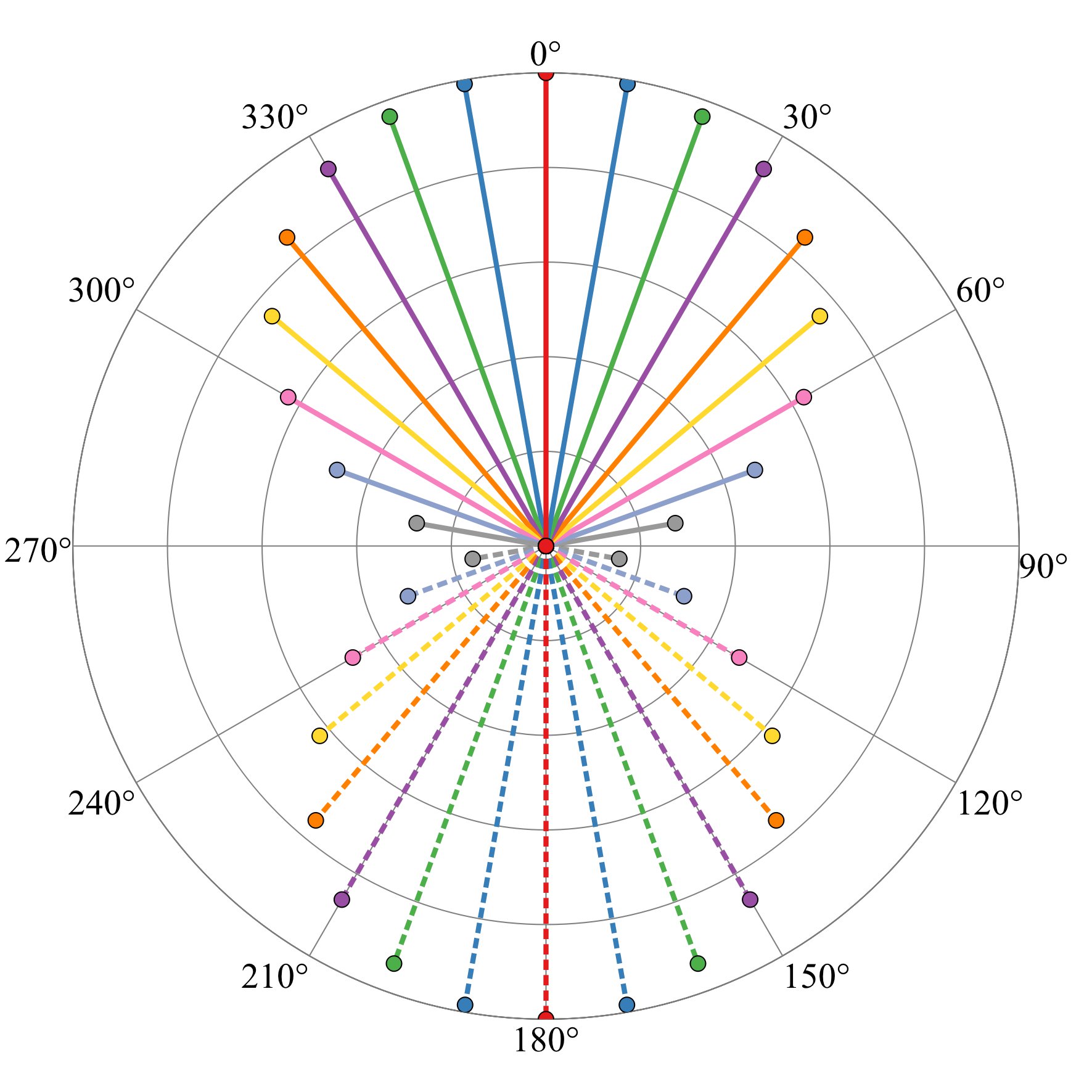}\par 
    \includegraphics[width=\linewidth]{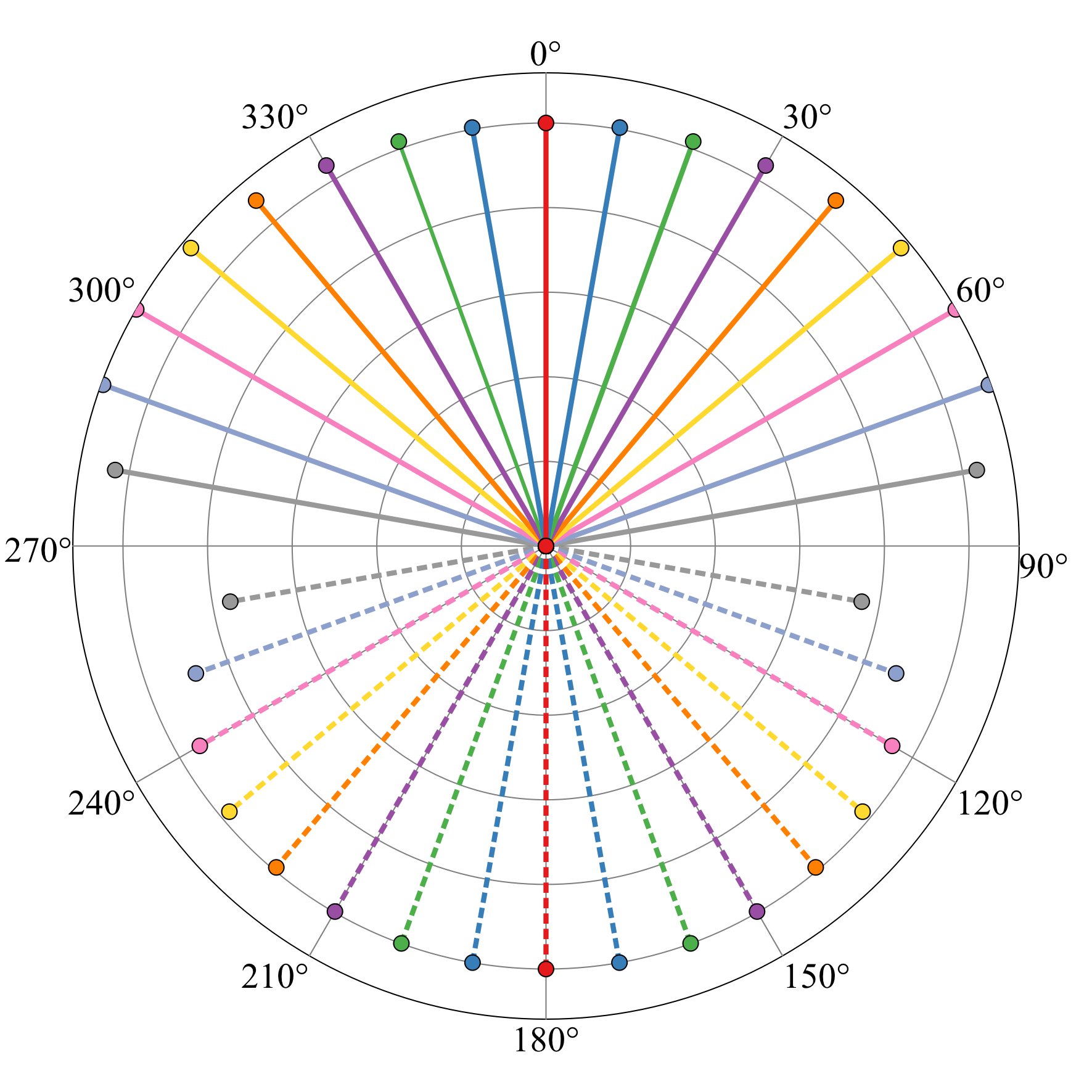}\par 
    \includegraphics[width=\linewidth]{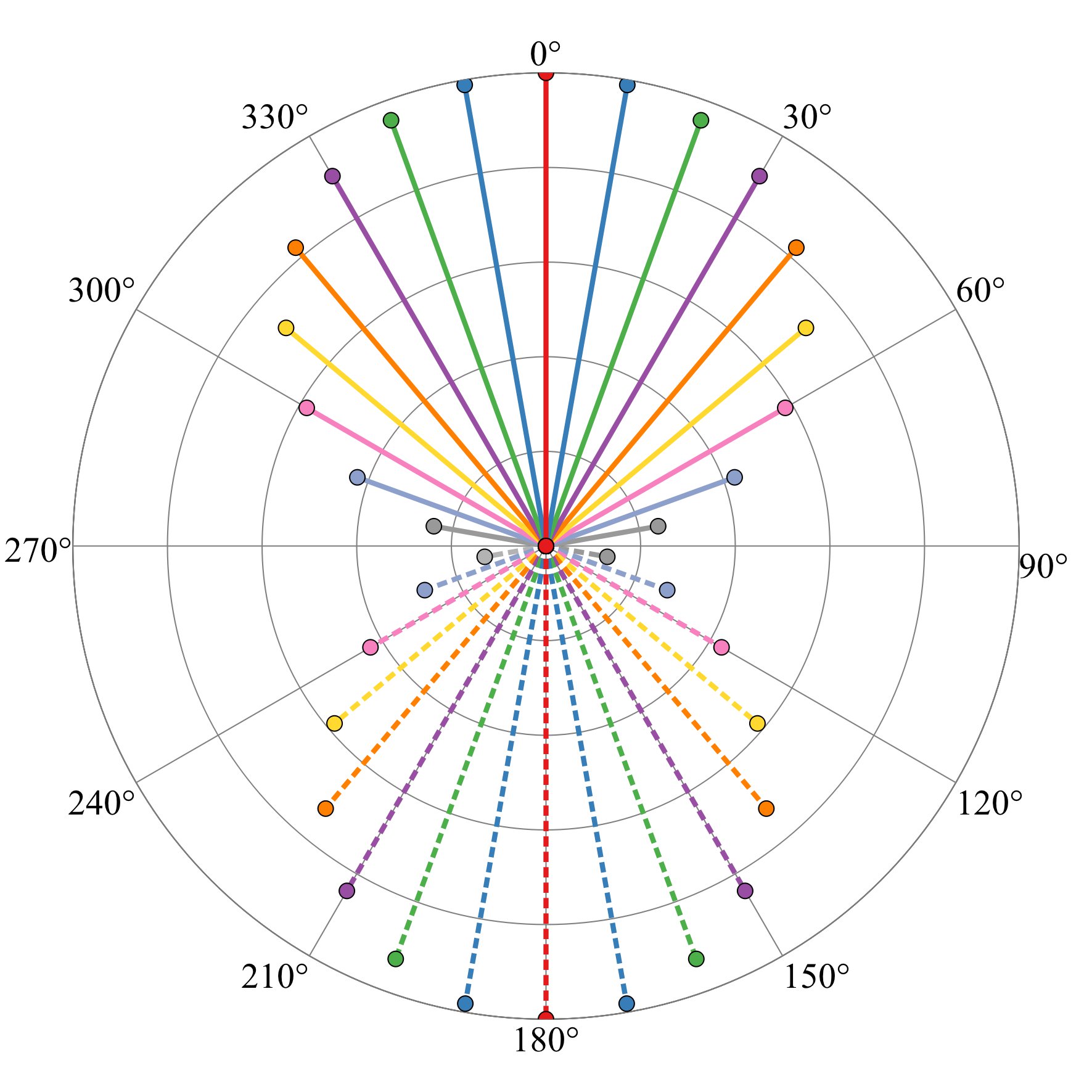}\par
    \end{multicols}
\caption{Polar diagrams of the radiation pattern for zero spin ($a = 0$, left panel), maximal spin ($a = 0.998$, middle panel), and retrograde spin ($a = -1$, right panel). The radiative flux values are normalised to the pole-on value, with the upper halves representing the case without limb-darkening (solid lines), while the lower halves include the limb-darkening effect (dotted lines). }
\label{Fig_polar_diagram}
\end{figure*}

Astrophysical black holes are uniquely characterised by two fundamental parameters: mass and spin. 
Coupled with the accretion rate, these parameters determine the variety of phenomena observed in active galactic nuclei (AGN). The event horizon of a BH is defined by the Schwarzschild radius, $R_S = 2 R_G = \frac{2 G M_{BH}}{c^2} \sim 3 \times 10^{13} M_8$ cm (for a non-rotating $10^8$-solar-mass black hole $M_{BH} = M_8 \times 10^8 M_{\odot}$), while a typical galactic radius is of the order of $R_{gal} \sim 1 \mathrm{kpc} \sim 3 \times 10^{21}$ cm. Thus there is a huge difference in physical scales (spanning $\sim$8 orders of magnitude) between the central BH and its host galaxy. 

The physical link connecting the two extreme scales is usually attributed to some form of AGN feedback \citep[][and references therein]{Silk_Rees_1998, Fabian_1999, King_2003, Murray_et_2005, Fabian_2012}. 
There is increasing observational evidence of such AGN feedback in action, detected in the form of powerful outflows on galactic scales \citep[e.g. see][for recent compilations]{Fiore_et_2017, Fluetsch_et_2019}. 
Galactic outflows can be driven by different physical mechanisms, such as radio jets and BAL winds, which may prevail at different stages in galaxy evolution (e.g. depending on the underlying accretion state). 
Here we focus on feedback due to outflows driven by AGN radiation pressure, expected when the accretion rate exceeds a few $\%$ of the Eddington limit.
We have previously discussed how such AGN feedback, driven by radiation pressure on dust, can account for the observed dynamics and energetics of galactic outflows \citep{Ishibashi_Fabian_2015, Ishibashi_et_2018}. 
For simplicity, we have up to now assumed the idealised case of spherical symmetry and isotropic emission.
However, in realistic situations, the optical/UV radiation should be  significantly anisotropic, as a result of emission from the accretion disc. 

Here, we analyse the angular dependence of the radiation pattern as a function of the BH spin, and its effects on the radiation-driven outflows on galactic scales. We recall that the dimensionless spin parameter defined as $a = cJ/GM^2$ (where $J$ is the angular momentum of the BH) varies from $a = 0$ for a non-rotating Schwarzschild BH to $a = 1$ for a maximally rotating Kerr BH. The associated innermost stable circular orbit (ISCO) is located at $r = 6 R_G$ for a Schwarzschild BH and $r = 1 R_G$ for a Kerr BH, respectively. As the last stable orbit is located at a smaller radius for higher BH spins, the radiative efficiency is higher (with a maximal value of the order of $\sim 30 \%$, taking into account the counteracting radiation torque, \citet{Thorne_1974}), and the relativistic effects are much stronger around rapidly spinning BHs \citep{Cunningham_1975}. The actual emission structure is thus determined by the BH spin, and the resulting radiation pattern is expected to be different for low-spin and high-spin objects. This difference should subsequently reflect on the large-scale morphology of the radiation-driven outflows. 

Therefore the spin of the central BH can have macroscopic effects, which directly imprint on the outflow geometry on galactic scales. This suggests that we can potentially deduce the BH spin from the observed shape of galactic outflows. In principle, this would provide a new way of constraining the central BH spin, provided that a precise mapping between radiation pattern and outflow geometry can be achieved. 
In the following, we explore this interesting possibility, outlining the basic concepts and discussing the physical implications in the context of AGN-galaxy co-evolution. 


\section{Radiation pattern}

The optical/UV radiation emerging from the accretion disc is expected to be anisotropic: maximal along the polar axis, and declining with increasing  inclination angle. According to the classical analytic approximation, the radiative flux follows a relation of the form: $\mathrm{Flux}_{rad}(\theta) \propto \cos \theta (1 + 2 \cos \theta)$, which takes into account the change in the projected surface area and the limb darkening effect \citep[e.g.][and references therein]{Kawaguchi_Mori_2010}. 
The effect of such anisotropic radiation on the distribution of surrounding dusty gas, and its impact on AGN obscuration, was previously considered by \citet{Liu_Zhang_2011}; but the relativistic effects due to the BH spin dependence were ignored in their work. The exact radiation pattern is determined by the central BH spin, with relativistic effects becoming extremely important for rapidly spinning BHs \citep{Sun_Malkan_1989, Li_et_2005}. In particular, the anisotropy can be reduced, e.g. due to gravitational focusing and light bending effects, which tend to redirect the radiation back to the disc. We thus need a precise computation, including all relativistic effects, in order to obtain the radiation pattern around BHs of different spins. 


\subsection{General relativistic accretion disc emission}
\label{Sect_KERRBB}

\begin{figure}
\begin{center}
\includegraphics[angle=0,width=0.4\textwidth]{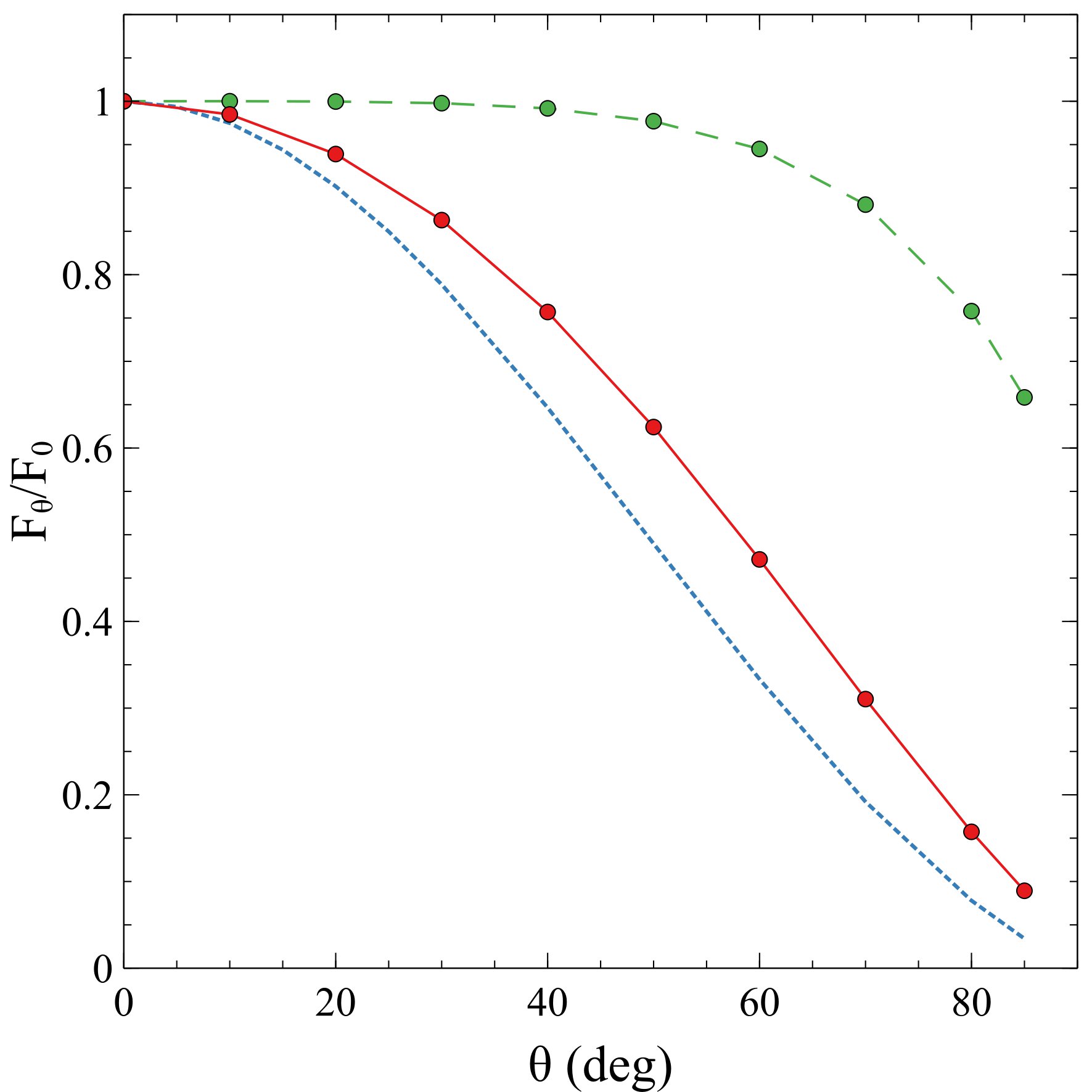}
\caption{ 
Radiative flux normalised to the pole-on value as a function of the inclination angle: analytic approximation (blue dotted line), and numerical results obtained with KERRBB for zero spin (red solid line) and maximal spin (green dashed line), including limb-darkening. 
}
\label{Fig_comp_analytic}
\end{center}
\end{figure} 

Here we compute the accurate angular dependence of the radiation pattern by using the KERRBB model implemented in XSPEC \citep{Li_et_2005}. KERRBB is a numerical code modelling a general relativistic accretion disc around a rotating BH, which takes into account all relativistic effects (such as Doppler boosting, gravitational redshift, and light bending), and includes the disc self-irradiation. 
We note that KERRBB was originally developed for fitting the spectra of stellar-mass black holes in X-ray binaries; here we are only interested in the angular dependence, and we apply the results to super-massive black holes by assuming scale-invariance. A note of caution should be added: the accretion disc is characterised by a temperature gradient decreasing outwards (with a radial profile of the form $T(r) \propto r^{-3/4}$ in the standard disc model), such that extreme UV photons originate from the innermost regions, while optical photons originate from outer radii. Nonetheless, our computation should still be valid, provided that the dusty outflows remain optically thick to the global disc radiation (a more detailed calculation may be required in future work, cf. Discussion). 

In Figure \ref{Fig_polar_diagram}, we show the resulting radiation pattern in the form of polar diagrams, for the case of zero spin (left panel), maximum spin (middle panel), and retrograde spin (right panel). 
In the  case of zero spin ($a = 0$), we see that the radiative flux is maximal along the polar axis ($\theta = 0^{\circ}$), and steadily decreases with increasing inclination angle. At $\theta = 85^{\circ}$, the flux is reduced by a factor of $\sim 6$ compared to the pole-on value. Thus the emission is vertically focused, peaking along the polar axis, and the resulting radiation pattern has a `prolate' form. 
In the case of maximum spin ($a = 0.998$), we observe that the radiation pattern is more isotropic, with the radiative flux varying much less with inclination angle (the greatest difference is only by a factor of $\sim 1.2$). 
In contrast to the case of zero spin, the maximal flux is not the pole-on value, but is instead reached at a larger angle ($\theta \sim 60^{\circ}$). The resulting radiation pattern has an `oblate' form. 
Comparing the two extreme cases, we note that the radiative flux can be much higher for maximal spin compared to zero spin, at all inclination angles: a factor of $\sim 3$ higher at $\theta \sim 0^{\circ}$, and up to a factor of $\sim 17$ higher at $\theta \sim 85^{\circ}$. 
We note that the polar diagram for retrograde spin ($a = -1$) is qualitatively similar to the case of zero spin, with the radiation pattern displaying a slightly more pronounced prolate form.

We further consider the effects of limb-darkening on the radiation pattern (shown as dotted lines in the lower halves of the polar diagrams in Fig. \ref{Fig_polar_diagram}). By including limb-darkening, the flux variations with inclination angle become more prominent. For instance, in the case of zero spin, the radiative flux at $\theta = 85^{\circ}$ is reduced by a factor of $\sim 11$ compared to the pole-on value, resulting in an even more vertically collimated emission pattern. A qualitatively similar argument may also apply for the cases of maximum spin (although the flux variations are much less pronounced) and retrograde spin. 

In Figure \ref{Fig_comp_analytic}, we show the comparison between the numerical results obtained with the KERRBB code and the classical analytic approximation. We note that the analytic relation may be a viable approximation for the zero spin configuration (at small inclination angles); but is clearly inadequate for high BH spins, where numerical results taking into account the full relativistic effects need to be considered. 


\section{Outflow geometry}
\label{Sect_Outflow_geoemtry}

We consider outflows driven by radiation pressure on dust \citep[cf.][]{Ishibashi_Fabian_2015, Ishibashi_et_2018}. Radiation originating from the accretion disc is absorbed by dust grains embedded in the gas, such that the surrounding dusty gas is swept up into an outflowing shell. The subsequent evolution is governed by the competition between the outward force due to radiation pressure and the inward force due to gravity.  Since the driving radiation field is anisotropic, with the radiative flux depending on the inclination angle, the effective Eddington ratio will also be a function of the inclination angle. As a result, the radiation pressure-driven outflows are expected to display an anisotropic configuration on galactic scales. Thus the large-scale geometry of the outflow depends on the underlying radiation pattern, which is in turn determined by the central BH spin. 

As we have seen in the previous section, the radiation field presents distinct patterns, depending on the spin of the central BH. For zero spin, the emission pattern is vertically focused, leading to `polar outflows' elongated along the polar axis. In contrast, for maximum spin, the radiation pattern has an oblate form, giving rise to `oblate outflows'. 
The global shape of the outflows can thus be significantly different for low-spin and high-spin sources, such that observations of the outflow geometry can already give us a hint of the central BH spin. 

\begin{figure}
\begin{center}
\includegraphics[angle=0,width=0.4\textwidth]{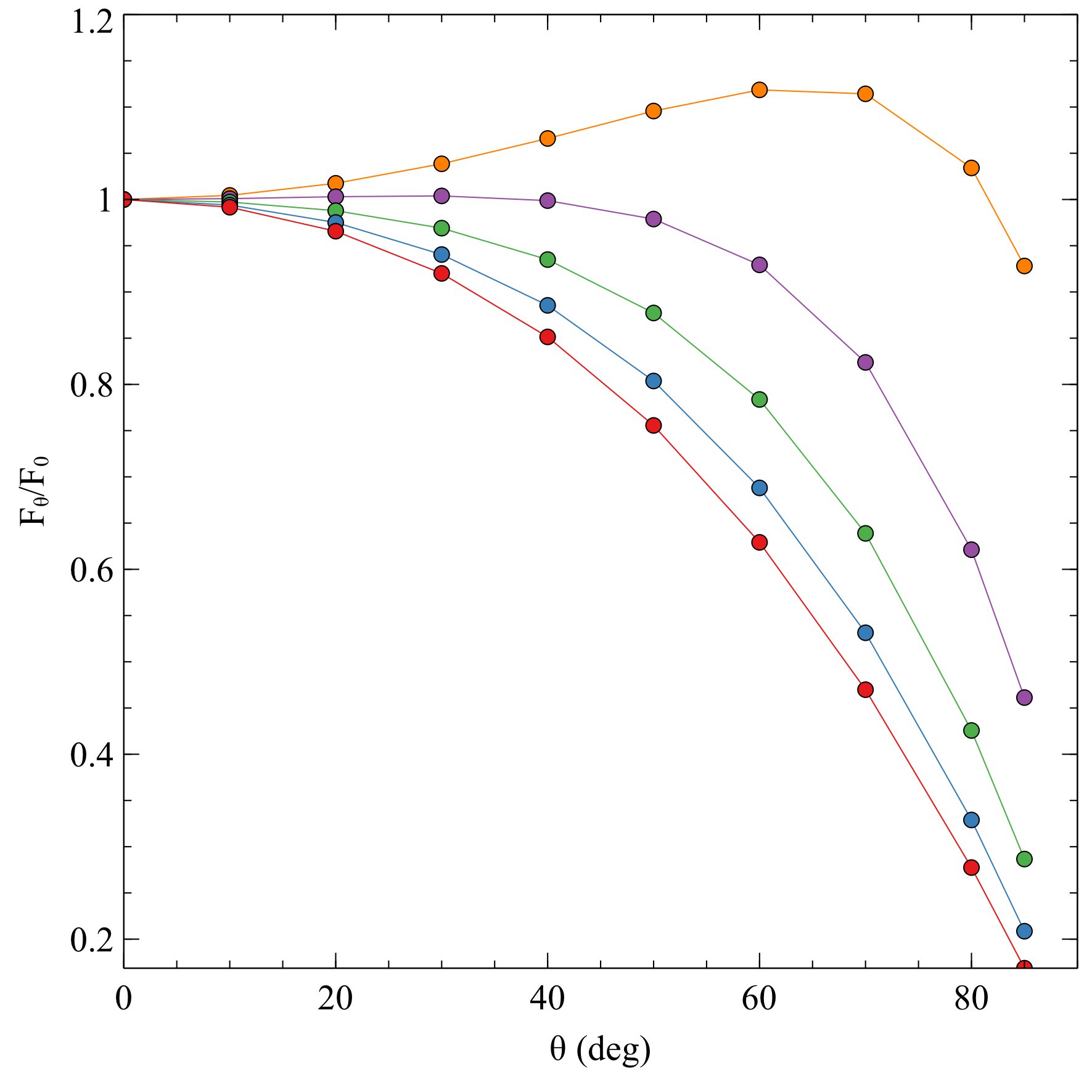} 
\caption{Radiative flux (normalised to the pole-on value) as a function of the inclination angle for intermediate spin values (without including limb-darkening): $a = 0$ (red), $a = 0.5$ (blue), $a = 0.8$ (green), $a = 0.95$ (violet), $a = 0.998$ (orange).
}
\label{Fig_intermediate_spins}
\end{center}
\end{figure}

Figure \ref{Fig_intermediate_spins} shows the angular dependence of the radiation pattern for intermediate values of the BH spin. We observe that for low-to-moderate spins ($0 \lesssim a \lesssim 0.8$), the radiation is  preferentially focused along the polar axis, leading to polar or prolate outflows. For higher BH spins, the radiation pattern becomes more isotropic, resulting in more broad quasi-spherical outflows. 
Therefore, prolate outflows should be the common outcome for the majority of BH spins, while oblate outflows may only be expected for the most extreme spin values.  

Furthermore, assuming the idealised case of thin shells outflowing at constant velocity, a difference in the velocity profiles of the emission lines can also be observed for different BH spins. A broader velocity profile (with a larger $\Delta v$ spread) may be expected for high BH spins compared to a narrower profile for low BH spins, simply due to projection effects. The observed difference in the velocity profiles can then give us a further indication on the central BH spin. 
We note that in reality, the outflow velocity is unlikely to be constant with radius, and the interpretation of the resulting velocity profiles will be more subtle, nevertheless a difference should still be discernible.  

\section{The $N_H - \lambda$ plane and obscuration}

\begin{figure*}
\begin{multicols}{2}
    \includegraphics[width=0.93\linewidth]{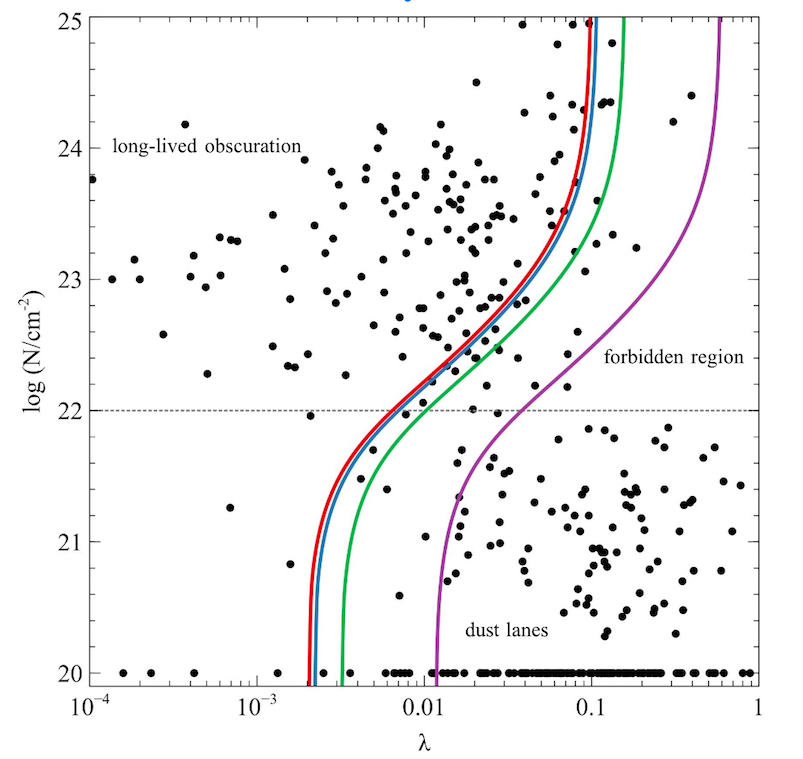}\par
    \includegraphics[width=0.93\linewidth]{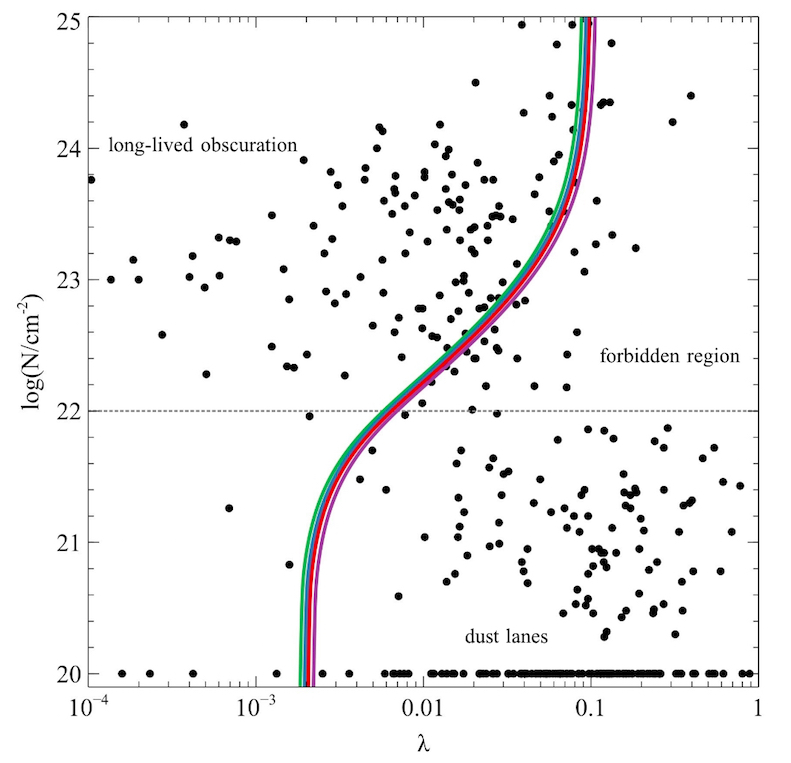}\par 
    \end{multicols}
\caption{
$N_H - \lambda$ plane in the case of zero spin ($a = 0$, left panel) and maximal spin ($a = 0.998$, right panel), for different inclination angles: $\theta = 0^{\circ}$ (red),  $\theta = 30^{\circ}$ (blue),  $\theta = 60^{\circ}$ (green),  $\theta = 85^{\circ}$ (violet). Observational data points from the Swift/BAT AGN sample, based on the 70 month Swift/BAT AGN catalogue \citep{Ricci_et_2017}. Unobscured sources are assigned an upper limit of logN
= 20. The horizontal line (grey fine-dotted) marks the $N = 10^{22} \mathrm{cm^{-2}}$ limit, below which absorption by outer dust lanes may become important. 
 }
\label{Fig_NH_lambda}
\end{figure*}

The actual importance of radiative feedback can be observationally tested by analysing how different AGN samples populate the `$N_H - \lambda$ plane', defined by the column density versus the Eddington ratio \citep[e.g.][]{Fabian_et_2008}. The effective Eddington limit for dusty gas defines two distinct areas in this plane: the region of `long-lived obscuration' and the so-called `forbidden' region. Observational results based on hard X-ray selected AGN samples, indicate that the effective Eddington limit is generally respected, with most sources avoiding the forbidden region, where they should be outflowing \citep{Fabian_et_2009, Raimundo_et_2010, Ricci_et_2017}. 
We have recently shown how the shape of the forbidden region is modified in the presence of radiation trapping \citep{Ishibashi_et_2018b}. 

As mentioned in Section \ref{Sect_Outflow_geoemtry}, due to the angular dependence of the radiation pattern, the effective Eddington ratio has an angular dependence, and so does the corresponding critical column density \citep[cf.][]{Ishibashi_et_2018}. In Figure \ref{Fig_NH_lambda}, we plot the resulting $N_H - \lambda$ plane for the case of zero spin (left panel) and maximum spin (right panel), with the data from the Swift/BAT AGN sample. We recall that the observational data are obtained for a given line of sight, hence the location of the sources in the $N_H - \lambda$ plane (either in the forbidden region or long-lived obscuration region) corresponds to that particular sightline. 

Dusty gas could be outflowing in one direction but not in others: the dust can be blown out in the face-on direction, but survive in the edge-on direction \citep[cf.][]{Liu_Zhang_2011}. 
A given source may be located in the forbidden region for low inclination angles (polar outflows), whereas it may be located in the region of long-lived obscuration for higher inclination angles. In contrast, in the case of maximum spin, the radiation pattern is much more isotropic, and the effective Eddington limit roughly converges for different inclination angles. For such maximally spinning BHs, there could be gas outflowing even at high inclination angles (quasi-spherical outflows), with the corresponding sources being located in the forbidden region. 

Overall, dusty gas can be easily blown out along the polar axis, clearing out the obscuration in the polar directions. On the other hand, the dusty gas can still survive close to the equatorial plane, leading to long-lived obscuration. Therefore, the obscuration geometry could be essentially determined by the radiation pattern, i.e. by the BH spin. 

We assume that the Eddington fraction is deduced via an isotropic indicator, such as the X-ray coronal emission. If however the corona is located close to the centre of the disk and corotates with it then the emission may be anisotropic and the situation more complicated than considered here. 


\section{Discussion}

\subsection{Potentials and caveats}

We have seen that the BH spin determines the emerging radiation pattern, which directly imprints on the large-scale geometry of the radiation pressure-driven outflows. Reversing the argument, the observed shape of galactic outflows may give us an indication on the underlying radiation pattern, hence the spin of the central BH. In principle, this may provide a novel way of probing the BH spin. In fact, despite the $\sim 8$ orders of magnitude difference in physical scales, the BH spin can have a major effect on the macroscopic properties observed on galactic scales. This suggests a new means of constraining the central BH spin from galactic-scale observations, provided that a precise mapping can be obtained. 

At present, the observational evidence of such BH spin-outflow connection may be unclear for individual objects, and we may not be able to deduce the spin for a particular source. However, this should become possible for larger samples that will be available in the near future, e.g. with eROSITA.  

Qualitatively, the observed morphology of the radiation-driven outflows may already provide a hint for the underlying BH spin. We predict that polar or prolate outflows are indicative of low-to-moderate BH spins, while quasi-spherical outflows are most likely associated with high BH spins. In particular, oblate outflows point towards the  most extreme spin values. Moreover, differences in the observed velocity profiles can also provide additional clues on the nature of the central BH spin. 

However, we are aware that the situation is much more complicated in reality, with several important caveats to keep in mind. 
For instance, the above arguments may be most relevant in the case of smooth spheroidal/elliptical galaxies, where the large-scale outflow geometry simply reflects the underlying radiation pattern. But this will not be the case for disc/spiral galaxies, where the outflow geometry may instead be shaped by the presence of the large-scale stellar disc. In the latter case, polar outflows may arise due to collimation by the galactic disc, rather than the underlying radiation pattern, and it may be difficult to disentangle the two effects. 
In addition, the actual gas distribution in galaxies is likely to be inhomogeneous and clumpy, further complicating the general interpretation. 
Overall, our results on the BH spin-outflow connection might be clearest for luminous quasars in elliptical galaxies; but, depending on the relative inclination of the Galactic and accretion discs,Ê may still have an effect on Seyfert nuclei, which tend to reside in disc galaxies where much of the gas is centrifugally supported. 

We further note that we are not accounting for any effects related to the finite geometric thickness of the accretion disk \citep{Taylor_Reynolds_2018a, Taylor_Reynolds_2018b}. These are expected to be modest for accretion rates below $10\%$, but may lead to an additional polar collimation of the radiation pattern (and hence under-estimation of black hole spin) at higher accretion rates.


\subsection{Polar outflows and obscuration geometry}

Recent interferometric observations have revealed that the bulk of the mid-IR emission originates from the polar regions, rather than from the equatorial dusty torus. Indeed, polar mid-IR emission has been detected in a number of local Seyfert galaxies, including type-1 AGN, extending on scales from a few parsecs up to hundreds of parsecs \citep{Asmus_et_2016, Lopez-Gonzaga_et_2016, Leftley_et_2018}. 

Such polar dust emission is not expected in classical unification scenarios \citep{Antonucci_1993}, and the new IR observations suggest a revision of the standard model. In this context, a two-component structure consisting of an inflowing disc and an elongated wind has been proposed \citep[][and references therein]{Hoenig_Kishimoto_2017}, and a similar geometric setup has also been applied to the particular case of the Circinus galaxy \citep{Stalevski_et_2017}. 
In our picture, elongated outflows tend to naturally develop along the polar axis, with the actual shape depending on the BH spin. In particular, we expect polar outflows for the majority of low-to-modest BH spins, which can easily account for the observed polar dust emission. 

Furthermore, the angular dependence of the radiation pattern may provide a natural physical interpretation for the obscuration geometry: dusty gas can be easily cleared out in the face-on direction by polar outflows, while it may survive at higher inclination angles. For rapidly spinning BHs, quasi-spherical outflows may efficiently remove dusty gas from most directions, except in the equatorial plane, where obscuration will be long-lived. 
Therefore even the obscuration geometry could be shaped by the radiation pattern, hence set by the BH spin. 
 

\subsection{Further implications}
\label{Sect_implications}

As we have seen, the central BH spin can have significant macroscopic effects on galactic scales. Physically, this can be related to two facts: a higher BH spin implies a smaller ISCO, whereby the radiative efficiency is higher and the relativistic effects are stronger. Thus the BH spin may also influence the strength of the outflow, in addition to its shape. In fact, the radiative flux can be much higher for maximally spinning BHs compared to zero spin objects (cf. Sect. \ref{Sect_KERRBB}). 
The greater driving force of rapidly spinning BHs imply higher speed outflows on larger scales.  
Coupled with the angular dependence, a high BH spin thus leads to powerful quasi-spherical outflows, which rapidly clear out the surrounding dusty gas. 

This suggests interesting implications for the co-evolutionary sequence, relating dust-obscured starbursts to unobscured luminous quasars \citep{Sanders_et_1988}. The obscuring dusty gas may be swiftly removed around rapidly spinning BHs, whereas the obscuration could be long-lived for low BH spins. 
As a consequence, the typical duration of the obscuration may depend on the central BH spin, with the obscured phase lasting longer/shorter for low/high spin. 
Such a spin-dependence may be included in the radiation-driven dusty shell models \citep{Ishibashi_et_2017} applied to the different populations of dusty quasars recently uncovered \citep{Banerji_et_2015, Zakamska_et_2016}.

The BH spin distribution can entail further effects. High-spin objects are preferentially detected in observational samples, due to their higher radiative efficiencies, which cause severe selection bias in flux-limited surveys \citep{Vasudevan_et_2016}. In addition, we expect a selection effect due to the angular dependence of the radiation pattern itself: for low-to-moderate BH spins, the emerging radiation is significantly reduced at large inclination angles, which can lead to a selection bias against high-inclination sources.  

One should also consider some important effects related to the spectral energy distribution. 
For instance, a harder UV spectrum is expected for a higher BH spin, while the peak emission is shifted towards lower energies for a larger BH mass. The precise outcome, including the mass and spin dependence, as well as the wavelength-dependent dust opacity, needs to be investigated through detailed radiative transfer calculations (e.g. CLOUDY).   

In conclusion, the BH spin can have a considerable impact on large scales, shaping AGN feedback and obscuration. Provided that a precise mapping can be obtained, we can in principle deduce the BH spin from the observed morphology of radiation-driven outflows. Alongside direct spin measurements, this may provide a novel way of probing the central BH spin.


\section*{Acknowledgements }

ACF and WI acknowledge ERC Advanced Grant 340442.
WI acknowledges support from the University of Zurich. 
CSR thanks the UK Science and Technology Facilities Council for support under Consolidated Grant ST/R000867/1.

  
\bibliographystyle{mn2e}
\bibliography{biblio.bib}


\label{lastpage}

\end{document}